\documentclass[aps,pra,twocolumn,showpacs,superscriptaddress]{revtex4}
\usepackage{graphicx}

\begin{document}


\title{Unexpected Magnetism of Small Silver Clusters}


\author{M. Pereiro}
\email[Email address: ]{fampl@usc.es}
\affiliation{Departamento de F\'{\i}sica Aplicada, Universidade de Santiago
de Compostela, Santiago de Compostela E-15782, Spain.}
\affiliation{Instituto de Investigaci\'ons Tecnol\'oxicas, Universidade de Santiago
de Compostela, Santiago de Compostela E-15782, Spain.}
\author{D. Baldomir}
\affiliation{Departamento de F\'{\i}sica Aplicada, Universidade de Santiago
de Compostela, Santiago de Compostela E-15782, Spain.}
\affiliation{Instituto de Investigaci\'ons Tecnol\'oxicas, Universidade de Santiago
de Compostela, Santiago de Compostela E-15782, Spain.}
\author{J. E. Arias}
\affiliation{Instituto de Investigaci\'ons Tecnol\'oxicas, Universidade de Santiago
de Compostela, Santiago de Compostela E-15782, Spain.}


\date{\today}

\begin{abstract}
The ground-state electronic, structural, and magnetic properties of small silver clusters, Ag$_n$
(2$\le$n$\le$22), have been studied using a linear combination of atomic Gaussian-type orbitals
within the density functional theory. The results show that the silver atoms,
which are diamagnetic in bulk environment, can be magnetic when they are grouped together in clusters. 
The Ag$_{13}$ cluster with icosahedral symmetry has the highest magnetic moment per atom among the studied
silver clusters. The cluster symmetry and the reduced coordination number specific of small clusters
reveal as a fundamental factor for the onset of the magnetism. 
\end{abstract}


\maketitle

In the last two decades, the research field of clusters has shown a rapid
development in both experimental and theoretical investigations \cite{haberland},
since the clusters are well suited for several applications. For example, there has
been a traditional interest in applications to catalysis \cite{henry}, due to
the considerable surface/volume ratio of clusters. More recently, 
clusters or nanoparticles that possess magnetic properties have offered exciting
new opportunities for biomedical applications including (i) magnetic 
separation of labeled cells; (ii) therapeutic drug
delivery; (iii) hyperthermic treatment for malignant cells; (iv) contrast enhancement 
agents for magnetic resonance imaging applications; (v) and also very recently for
manipulating cell membranes \cite{berry}.

Clusters are on the border line between atoms and bulk and thereby, they play
an important role in understanding the transition from the microscopic structure to the
macroscopic structure of matter. Although 4$d$ and 5$d$ transition metal atoms have unfilled
localized $d$ states, none of them are magnetic. Only a few of the 3$d$ transition metals 
form magnetic solids. Thus, from the magnetic point of view, one of the long
standing problems in condensed matter physics is to understand why some
nonmagnetic metals become magnetic when they condense into clusters. There are two factors
characteristic of clusters that mainly contribute to the onset and enhancement of the magnetism,
namely, the reduced coordination number and the high symmetry since that, symmetry enables degeneracy and 
degeneracy spawns magnetism \cite{dunlap}. According to this, an icosahedral structure is a good
candidate for the appearance of the magnetism because the maximal degeneracy of an irreducible
representation of the icosahedra (I$_h$) group is 5 whereas all other cluster symmetries allow
at most three-fold degeneracy, as was reported by Reddy {\it et al.} for Pd, Rh, and Ru \cite{reddy}. 

In this letter, we present first-principles calculations on small silver clusters
exhibiting an important magnetism,
which has not been predicted until now.
In the present work, we study the evolution of the magnetism of silver clusters as a function of the
cluster size and special emphasis is placed on the effects of cluster symmetry. 
The magnetic behavior that we got is very interesting, if not astonishing, compared
with the silver bulk magnetic properties. To the best knowledge of us, nobody has predicted or
even observed magnetism in small bare silver clusters, however our computational simulations anticipate
the onset and enhancement of the magnetism for the silver clusters that gather special conditions 
like high symmetry, and reduced coordination number.  In this respect,
the Ag$_{13}$ cluster exhibits the highest magnetic moment among the studied clusters and converges into
a very stable 
structure. 

\begin{figure}
\includegraphics[width=8.5cm,angle=0]{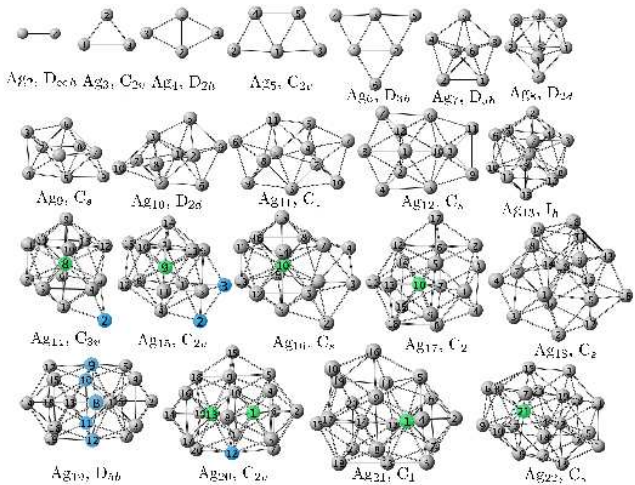}
\caption{\label{fig1}(Color online). Lowest energy structures and the symmetry point groups 
assigned to the Ag$_n$ (2$\le$n$\le$22)
clusters. The magnetic moments, namely m, of the atoms showed in green 
(-1.5~$\mu_B$$\le$m$\le$-0.4~$\mu_B$) and blue
(-0.1~$\mu_B$$\le$m$\le$0~$\mu_B$) are aligned antiferromagnetically to the other ones.}
\end{figure}

With the aim of studying the magnetic properties of small silver clusters,
Ag$_n$ (2$\le$n$\le$22), we have
performed density functional theory-based calculations consisting of 
a linear combination of Gaussian-type-orbitals Kohn-Sham density-functional
methodology as implemented in the {\sc demon-ks3p5} program package \cite{salahub}. 
All-electron spin-unrestricted calculations were carried out
at the generalized gradient approximation (GGA) level to 
take the exchange-correlation (XC) effects into account \cite{perdew}. 
Local-density approximation 
sometimes yields inaccurate bond lengths and total energies
due to the insufficiency in describing the strong correlation effect of the
localized $d$ electrons and charge density inhomogeneities. In these regards, the
GGA should be a better choice \cite{pereiro}.
For this reason, at the beginning of this work and to satisfy ourselves that the 
numerical procedure is reliable, we initiate a search of the functional that better
fitted the calculated bond length of the silver dimer to the experimental one. We found
that the functional that better fitted the bond length was the one developed by
Perdew and Wang \cite{perdew}, given a bond length of 2.535~\AA, that is in excellent 
agreement with the experimental measure (2.53350~\AA) reported in Ref.~\cite{simard}.
An orbital basis set of contraction pattern (633321/53211*/531+) was used in conjunction
with the corresponding (5,5;5,5) auxiliary basis set for 
describing the $s$, $p$ and $d$ orbitals \cite{huzinaga}. The grid for numerical
evaluation of the XC terms had 128 radial shells of points and each shell had
26 angular points. Spurious one-center contributions to the XC forces, typically 
found in systems with metal-metal bonds when
using a nonlocal functional, are 
eliminated in a similar way as has been done in Ref.~\cite{versluis}.
A wide set of spin multiplicities ranging from 1 to 11 was checked to
ensure that the lowest-energy electronic and magnetic configuration is reached.
The geometries were fully optimized without symmetry and
geometry constraints using the Broyden-Fletcher-Goldfarb-Shanno
algorithm \cite{broyden}. During the optimization, the convergence criterion 
for the norm of the energy gradient was fixed to $10^{-4}$ a.u. while it was $10^{-7}$ a.u. for 
the energy and $10^{-6}$ a.u. for the charge density.
A huge sampling of trial geometries taken from the literature was evaluated. 
While for these small clusters it is nearly impossible to search for all
possible geometries, the detailed search that we have carried out give us some confidence that
the structural minima has been found. For clusters
with size varying from 2 up to 12, the structural minima resulting from our calculations 
are closely related to the ones reported by
R. Fournier~\cite{rene},
whereas all other cluster structures are quite similar to the ones of Ref.~\cite{doye}. 
In the early stage of the geometry optimization process for silver clusters with size
ranging from 13 to 22 and with the aim of speeding up the calculations, 
the structure of the silver clusters was first optimized in conjunction
with a 17-electron scalar relativistic model core potential designed for the adequate
description of the silver dimer bond length \cite{andzelm}. Once the geometry of the 
cluster was converged for the model core potential, an all-electron
structural-relaxation calculation was performed leading to the current lowest energy 
structures showed in Fig.~\ref{fig1}.

\begin{figure}
\includegraphics[width=8.5cm,angle=0]{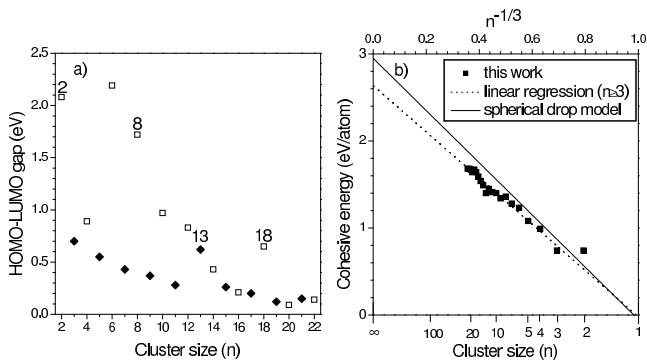}
\caption{\label{fig2}(a) The HOMO-LUMO gap, and (b) the cohesive energy 
per atom of small silver clusters as a 
function of the cluster size n (2$\le$n$\le$22).} 
\end{figure}

One of the criteria for a cluster to be used as a potential building block for a nanomaterial
is its chemical stability relative to other clusters of the same material. With the aim 
of studying the stability 
of Ag$_2$-Ag$_{22}$ clusters,  we have plotted the HOMO-LUMO gaps, and the cohesive energies 
in Fig.~\ref{fig2}(a),(b). The cohesive energies for Ag clusters are fitted
to a linear regression and compared in Fig.~\ref{fig2}(b) with the spherical droplet model of Miedema
\cite{miedema}
\begin{equation}
	E_c(n)=E_{bc}-\left(\frac{36\pi}{n}\right)^{1/3}\gamma^0 {V_a}^{2/3},
\end{equation}
where E$_{bc}$ is the bulk cohesive energy, $\gamma^0=7.8\times10^{18}$~eV/m$^2$ is the surface energy
of the bulk silver, V$_a$ is the atomic volume, and n is the number of atoms in the cluster. Despite
that some clusters (2$\le$n$\le$6) are far from spherical, our calculated cohesive energies are in good
agreement with the Miedema's model. Indeed, the bulk cohesive energy predicted by our calculations
(2.63 eV/atom) differs by only 10\%~ from the experimental findings (2.95 eV/atom) \cite{kittel},
given us confidence that the lowest energy structures plotted in Fig.~\ref{fig1} are reliable. In
Fig.\ref{fig2}(a) we can see that the most stable even-numbered clusters are Ag$_2$, Ag$_6$, Ag$_8$
and Ag$_{18}$. Except for Ag$_6$, it is a consequence of the closure of the electronic shell 
\cite{pereiro1}. For the odd-numbered clusters, the HOMO-LUMO gaps decrease monotonically as the cluster size
increase, except for the Ag$_{13}$ cluster. The large HOMO-LUMO gap of Ag$_{13}$ relative to the 
odd-numbered clusters
enhances its chemical stability and inertness, as well as its ability to assemble into magnetic nanoparticles
because it is a magnetic cluster with a considerable magnetic moment per
atom, as is discussed below. 

\begin{figure}
\includegraphics[width=8.5cm,angle=0]{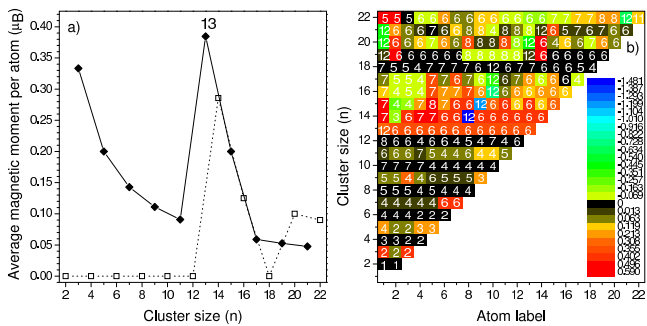}
\caption{\label{fig3}(Color online). Evolution of magnetic moment per atom of the most stable cluster isomers 
of each size. (a) Average magnetic moment per atom versus cluster size. (b) Color representation of the
magnetic moment per atom of each silver cluster. The numbers inside the small charts denote the coordination
number. On the right side, the color palette gives information
about the numerical values of the atomic magnetic moments. The numbers labeling every atom 
of each cluster in Fig.~\ref{fig1} are closely related to the ones displayed in x-axis.} 
\end{figure}

Our results on the magnetic properties of the Ag$_n$ (2$\le$n$\le$22) clusters are most fascinating,
if not unexpected, compared with the silver bulk magnetic properties. Indeed, the silver bulk
magnetic ordering is well-known to be diamagnetic, as can be inferred from its negative
magnetic susceptibility ($\chi_m$=-19.5$\times$10$^{-6}$ cm$^3$/mol), whereas silver atoms when
they coagulate to a cluster become either magnetic or non-magnetic depending on the cluster size, as 
is shown in Fig.~\ref{fig3}. 
For Ag$_2$, Ag$_8$ and Ag$_{18}$ clusters, our calculations clearly show
that they are highly stable with large HOMO-LUMO gaps [see Fig.~\ref{fig2}(a)] and diamagnetic
[see Fig.~\ref{fig3}]. The phenomenon can be understood simple from the 2, 8, and 18-electron
rule, as described by the cluster shell models \cite{crabtree}. The former rule predicts transition
metal clusters to be stable and diamagnetic when the valence shell of the metal atom contains 2, 8 or 18
electrons that completely fill electron shells forming closed electronic structures with paired spins.

From Ag$_3$ up to Ag$_{12}$ excluding Ag$_2$ and Ag$_8$, the cluster topologies converge into a geometry where all atoms belong to
the surface [Fig.~\ref{fig1}], and consequently with a reduced coordination number [Fig.~\ref{fig3}(b)].
This condition favors a reduced charge accumulation in nearest neighbor atoms that according to 
the calculated Mulliken population analysis (MPA)
is less than 0.1 electron per atom in average. Likewise, the Mayer bond order analysis reveals that silver
atoms are sharing the unpaired 5$s$ electrons forming a covalent bond and thereby, low spin configurations
are expected. For example, the even-numbered clusters become diamagnetic because they have an even number 
of 5$s$ electrons that create a cloud of paired 5$s$ electrons [Fig.~\ref{fig3}(a)], whereas 
for the odd-numbered clusters the electronic configuration with one unpaired spin electron 
is favored energetically over the whole feasible spin configurations studied in this work. The resulting 
odd-numbered clusters retain an average magnetic moment per atom that decreases monotonically with 
the increase of the coordination number [Fig.~\ref{fig3}(a),(b)] 
since the orbital overlap expands 
as the coordination number does, such as was reported for Fe, Co and Ni in Ref.~\cite{liu} and 
for 13-atom clusters of Pd, Rh, and Ru in Ref.~\cite{reddy}. 

\begin{figure}
\includegraphics[width=8.5cm,angle=0]{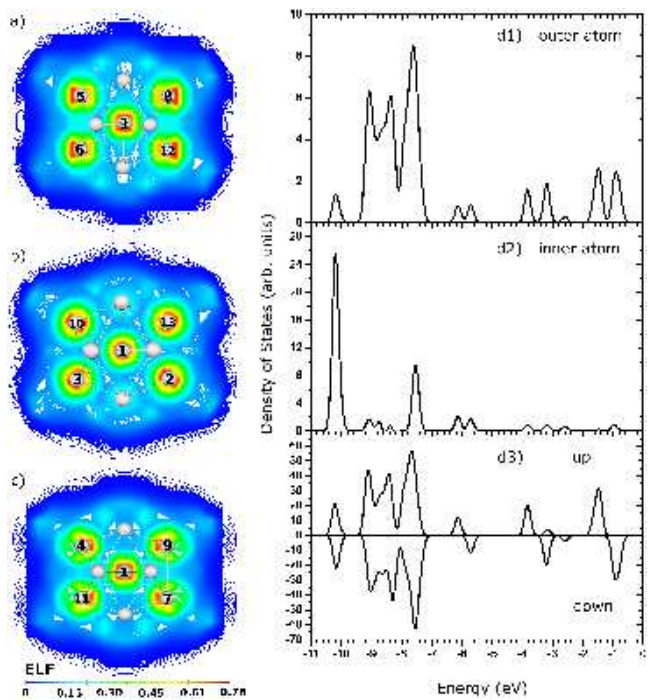}
\caption{\label{fig4}(Color online). Contour plot of the electron localization function for the three 
mutually orthogonal golden planes ((a), (b), (c)) of the icosahedral symmetry and densities of
states (DOS) for the outer-shell atoms (d1), and inner atom (d2) in an Ag$_{13}$ cluster. The spin-polarized
DOS is shown in (d3). The dashed vertical line represents the Fermi level.}
\end{figure}

In Fig.~\ref{fig3}(a) we can see that
Ag$_{13}$ exhibits the highest average magnetic moment per atom among the studied clusters. To
understand the origin of this considerable magnetic moment (0.39 $\mu_B$/atom) 
we show in Fig.~\ref{fig4}(a)-(d3)
the electron localization function (ELF) and the densities of states (DOS) for the inner and the outer-shell
atoms. The MPA reveals that the electron charge transfers from the outer-shell
Ag atoms to the central Ag atom which gains an excess of charge of about 0.8 electrons. The 
charge transfer, that is favored by the icosahedral symmetry, can be described graphically from our corresponding ELF plot. 
In Fig.~\ref{fig4}(a)-(c), the contour plot of the ELF indicates a slight concentration
of the light blue color (0.15-0.30 in bar color) surrounding the central atom, and give rise to
an enhancement of the electron localization around inner atom. The large coordination number of
the central Ag atom results in an enhancement of the overlap of its 4$d$ orbitals with those of other
outer-shell Ag atoms. Therefore, for the inner atom the charge transfer populate mainly
the lower energy 4$d$ states ($\sim$~-10.2 eV) with reduced exchange splitting whereas
for the outer-shell atoms these states become less occupied, as is shown
in Fig.~\ref{fig4}(d1),(d2) \cite{footnote} and Fig.~\ref{fig4}(d3), respectively. 
Thereby, the charge transfer lead to a small loss of the spin-up DOS at the Fermi level  for the inner atom
compared with the outer-shell atoms and 
consequently, the inner atom weakens its magnetism (0.35~$\mu_B$/atom) and the outer-shell
atoms enhance their magnetic moments (0.39~$\mu_B$/atom) given rise to an increase of the average magnetic
moment of the Ag$_{13}$ cluster. 

From Ag$_{14}$ to Ag$_{22}$ except for Ag$_{18}$ that have already been studied above, the clusters converge 
into a distorted icosahedral symmetry that depending on the cluster size, the number of inner atoms
is either one or two (see Fig.\ref{fig1}). In this case, the MPA confirms
a charge transfer mainly from the peripheral atoms to the inner ones even greater than the Ag$_{13}$ case.
This effect decompensate the spin pairing state and give rise to an antiferromagnetic alignment of the inner
atoms to the outer ones and thus, the clusters decrease the total magnetic moment compared with the Ag$_{13}$
cluster, as is shown in Fig.~\ref{fig1} and Fig.~\ref{fig3}(b). It is also observed in Fig.~\ref{fig3}(a)
that the average magnetic moment per atom decreases as the cluster size gets bigger up to n=19
and then, oscillates tending to decrease. The phenomenon can be
understood simply from the loss of symmetry that reduces orbital degeneracy and 
weakens magnetism \cite{dunlap}. According to this tendency, we expect for greater clusters in size 
an enhancement of the magnetic moment as long as the cluster stabilizes in a geometry of high symmetry (e.g.,
Ag$_{38}$, Ag$_{55}$, and Ag$_{75}$). 

In conclusion, we have shown that the silver atoms can be magnetic when they are grouped together in
small clusters. Particularly, the Ag$_{13}$ cluster exhibits the highest magnetic moment per atom
among the studied silver clusters due to its high symmetry and degeneration. 

The authors acknowledge the CESGA (Centro de Supercomputaci\'on de Galicia),
especially A. G\'omez and C. Fern\'andez for computational 
assistance. The work was supported by the Xunta de Galicia under
the Project No. PGIDIT02TMT20601PR.


\begin{thebibliography}{24}
\bibitem[{foo()}]{footnote}
\bibinfo{note}{In Fig.~\ref{fig4}(d1), we show the DOS for the atom labeled
  with the number 2 in Fig.~\ref{fig1} as a representative example of the DOS
  for the outer-shell atoms, since from symmetry considerations the external
  atoms have all of them similar DOS plot.}
\end{thebibliography}
\end{document}